\documentclass{emulateapj}
\usepackage{color}
\usepackage{latexsym, graphicx, amssymb, longtable,epsf,ulem}
\usepackage{url}
\usepackage{booktabs}
\usepackage{multirow}
\usepackage{txfonts}
\usepackage{CJK}
\usepackage{color}

\newcommand{\ym}{\color{black}}
\newcommand{\xie}{\color{black}}
\newcommand{\xiee}{\color{black}}

\long\def\symbolfootnote[#1]#2{\begingroup\def\thefootnote{\fnsymbol{footnote}}\footnote[#1]{#2}\endgroup}

\shorttitle{Photo-evaporation and Exomoon}
\shortauthors{Yang et al.}

\begin{document}

\title{Global Instability of Exo-Moon System Triggered by Photo-Evaporation}

\author{Ming Yang, Ji-Wei Xie$^{\ast}$, Ji-Lin Zhou$^{\ast}$, Hui-Gen Liu, Hui Zhang}
\affil{ School of Astronomy and Space Science \& Key Laboratory of Modern Astronomy and Astrophysics in Ministry of Education, Nanjing University, 210093, China}
\affil{$^{\ast}$Corresponding to jwxie@nju.edu.cn or zhoujl@nju.edu.cn}

\begin{abstract}
Many exoplanets have been found in orbits close to their host stars and thus they are subject to the effects of photo-evaporation. Previous studies have shown that a large portion of exoplanets detected by the {\it Kepler} mission have been significantly eroded by photo-evaporation. In this paper, we numerically study the effects of photo-evaporation on the orbital evolution of a hypothesized moon system around a planet. We find that photo-evaporation is crucial to the stability of the moon system.  Photo-evaporation can erode the atmosphere of the planet thus leading to significant mass loss. As the planet loses mass, its Hill radius shrinks and its moons increase their orbital semi-major axes and eccentricities. When some moons approach their critical semi-major axes, global instability of the moon system would be triggered, which usually ends up with two, one or even zero surviving moons. Some lost moons could escape from the moon system to become a new planet orbiting the star or run away further to become a free-floating object in the Galaxy. Given the destructive role of photo-evaporation, we speculate that  exomoons are less common for close-in planets ($<0.1$ AU),  especially those around M-type stars, because they are more X-ray luminous and thus enhancing photo-evaporation. The lessons we learn in this study may be helpful for the target selection of on-going/future exomoon searching programs.
\end{abstract}
\maketitle

\section{INTRODUCTION}
At the time of writing, the {\it Kepler} mission has detected more than 4600 planetary candidates, and over 1000 of them have been confirmed, boosting the total exoplanet number over 2000 (http://exoplanet.eu/). In contrast, detection of an exomoon still remains elusive \citep[and the references therein]{Kip14}, although moons are much more frequent than planets in our solar system. The non-detection could be because (1) technically, the signal of an exomoon is too weak to be captured \citep{Kip12} and (2) theoretically, few exomoons could exist around currently known exoplanets after a long formation and evolution process. 
Several effects which could affect the dynamical evolution of an exomoon have been studied in literature. 

First, from the view of dynamical stability, the orbit of a moon should be within the Hill radius of its host planet. Using detailed numerical simulations, \citet{Dom06} have shown that the critical semi-major axis of a moon can be expressed as 
\begin{equation}
a_{\rm cr} = \alpha  R_{\rm H}, 
\label{acr}
\end{equation}
where,
\begin{equation}
R_{\rm H} = a_{\rm p}\left(\frac{M_{\rm p}}{3M_*}\right)^{1/3}
\end{equation}
is the Hill radius,  $M_{\rm p}$ and $M_*$ are the masses of the planet and the central star, respectively, and $a_{\rm p}$ is the orbital semi-major axis of the planet. According to \citet{Dom06}:
\begin{eqnarray}
\alpha &\approx& 0.4895 (1-1.0305 e_{\rm p} - 0.2738 e_{\rm m}) \, \ \ \  {\rm prograde \ moons}, \nonumber \\
           &\approx& 0.9309 (1- 1.0764 e_{\rm p} - 0.9812 e_{\rm m}) \, \ \ \   {\rm  retrograde \ moons},
\label{alpha}
\end{eqnarray}
where $e_{\rm p}$ and $e_{\rm m}$ are the orbital eccentricities of planet and moon, respectively. Hereafter, subscript `p' and `m' denote properties of planet and moon, respectively. Note, orbital elements of planets are with respect to the central star, while those of moons are with respect to their host planets.   
Recently, using numerical integrations, \citet{Pay13} found $\alpha$ would be slightly smaller for tightly packed inner planets.

Second, tidal effects could expand or shrink the orbit of a moon, depending on a few initial parameters, such as the moon's orbital period and the planet's rotation period. In certain circumstances, the moon could hit on the planet's surface or escape from orbiting the planet\citep{BO02}. In some other circumstances, the moon could be evaporated or completely melted due to tidal heating \citep{Cas09}. However, one should note that the tidal effects have large uncertainty because of the poorly constrained tidal Q values of exoplanets and exomoons, which could vary in a range of several orders of magnitude.

Third, a close-in planet-moon system might not be formed in-situ, namely it could be formed {\ym in the outer part of} planetary disk and {\ym have} migrated inward to the current orbit via certain mechanisms, e.g., type I or II migration \citep[and the references therein]{PT06}. The inward migration would shrink the Hill radius of the planet and sometimes could induce resonances between moons and the planet, causing orbital instability and ejection of the moons \citep{Nam10, Spa16}. In addition, if planets underwent planet-planet scattering process, they are likely to have lost their moons \citep{Gon13}. 

For the thousands of planet candidates detected by {\it Kepler} mission, it has been suggested that the majority of them could be formed in-situ without large scale migration \citep{CL13, HM13}, although an opposite scenario has  been also proposed \citep{TP07, RC14}.  Most of these candidates, especially those in multiple transiting systems, are found to have low orbital eccentricities ($e_{\rm p}<0.1$) based on their transit durations \citep{VA15, Xie16} and timing variations \citep{WL13,HL14}, suggesting they are unlikely to have undergone planet-planet scattering{\xiee, unless their orbits were subsequently damped via certain mechanisms, e.g., tidal effects \citep{Fab07}}. Nevertheless, a lager portion of these candidates are thought to have undergone the photo-evaporation process \citep{OW13} due to their proximity to the central stars.  As photo-evaporation generally operates on a time scale {\xiee of $10^7-10^8$ yr generally} shorter than that {\xiee (on order of $10^9$ yr)} of tidal effects, it thus plays a crucial role in determining the path of the dynamical evolution of planet-moon systems. 

In this paper, we numerically investigate the effects of photo-evaporation on the dynamical evolution of planet-moon systems. 
In section 2, we describe our numerical model and present the results. Discussions and summary are in section 3 and 4.

\section{SIMULATIONS AND RESULTS}
We use the N-body simulation package --- MERCURY \citep{Cha99} --- to numerically investigate the effects of photo-evaporation on the dynamical evolution of planet-satellite systems. {\xie We choose the Bulirsch-Stoer integration algorithm, which can handle close encounter accurately. This is important in the simulations as we will see below that many close encounters among moons and the planet are expected to happen. Collisions among moons, the planet and the central star are also considered in simulations and treated simply as  inelastic collisions without fragmentations.} In each simulation, it consists of a central star, a planet and some moons orbiting around the planet. The photo-evaporation is simply  modeled as a slow (adiabatic) and isotropic mass loss process of the planet.  In reality, the photo-evaporation is a very slow process on a time scale of order of $10^{7}-10^{8}$ yr \citep{OW13}. 
However, it is impractical and unnecessary to perform a simulation on such a long time scale. Instead, we model the mass-loss process on a time scale of $\tau_{\rm evap}$, and each simulation typically lasts for several $\tau_{\rm evap}$. {\xie As long as the adiabatic requirement is met, i.e., mass-loss time scale is much longer than the dynamical time scale of the system ($\tau_{\rm evap} \gg P_{\rm p}$, where $P_{\rm p}$ is the orbital period of the planet),  one could study the dynamical effects of the mass-loss process equivalently. As we discussed in section 3.3, the results converge if $\tau_{\rm evap}>10^2-10^3$ $P_{\rm p}$, indicating the adiabatic condition is met. Therefore, in all other simulations we set $\tau_{\rm evap}=10^4$ $P_{\rm p}$.} 
{\xie Other parameters are set to represent the typical values of {\it Kepler} planets.} In particular, we consider a planet-satellite system orbiting a star of solar mass ($M_\star=M_\odot$) in a circular orbit ($e_{\rm p}=0.0$) with semi-major axis of $a_{\rm p}=0.1$ AU.  {\xie The orbit has a period of $\sim10$ days (typical value of Kepler planets), and it is sufficiently close to the central star to be subject to significant {\ym photo-evaporation} effect \citep{OW13},} {\ym which removes massive hydrogen envelopes of the planet.} The planet has an initial mass of $M_{\rm pi}$ and a final mass of $M_{\rm pf}$ after photo-evaporation. In this paper, we adopt $M_{\rm pi}=20 M_\oplus$ and $M_{\rm pf}=10 M_\oplus$ nominally (close to the standard model adopted in \citet{OW13}).  {\xie The mean density of the planet is set as the same to Neptune (1.66 $g/cm^3$). The effect of changing the planetary density is discussed in section 3.3.} We performed a number sets of simulations by considering different planet-satellite configurations.  Similar to the definition in MERCURY, hereafter, we define ``small moons" as test particles (TPs) whose mutual gravity and corresponding effects on the planet and the star are ignored, while ``big moons" {\ym are gravitationally important enough that their} gravitational effects are fully considered. Table \ref{tb_sim} lists the initial setups and parameters of various simulations, whose results are presented in the following subsections.

\subsection{Simulation A: all small moons}
We first consider the simplest case (Simulation A in Table \ref{tb_sim}), where a planet and its moons are initially in coplanar and circular orbits. The moons are treated as test particles (TPs), namely their mutual gravity and corresponding effects on the planet and the star are ignored. In this case, we aim to both analytically and numerically understand how the planet photo-evaporation (mass loss equivalently) process affects the orbital evolutions of its moons. 

Analytically, from the view of the planet-moon two-body problem, the adiabatic mass loss of the central body (i.e., the planet) causes orbital expansion of its orbiter, namely the increase of semi-major axis of the moon\citep{Had63}:
\begin{eqnarray}
a_{\rm m}(t) \approx \frac{M_{\rm p}(t_0)}{M_{\rm p}(t)} a_{\rm m}(t_0),
\label{a_th}
\end{eqnarray} 
where $t$ and $t_0$ denote the evolutional time and the initial time, respectively. As the moon's orbit expands, it will be subject to stronger tidal perturbation from the third body, i.e., the star, which increases the moon's orbital eccentricity \citep{Cas09} as
\begin{equation}
e_{\rm m} (t)\approx \left(\frac{P_{\rm m}}{P_{\rm p}}\right)^2 = \frac{M_*}{M_{\rm p}(t)} \left(\frac{a_{\rm m}(t)}{a_{\rm p}}\right)^3 = \frac{1}{3} \left(\frac{a_{\rm m}(t)}{R_{\rm H}(t)}\right)^3
\label{e_th}
\end{equation}
where $P_{\rm p}$ and $P_{\rm m}$ are orbital periods of the planet and the moon. Note, here the planet's Hill radius is a function of time; it shrinks as the planet loses mass due to photo-evaporation.

\begin{center}
\begin{figure}
\epsscale{0.65}
\plotone{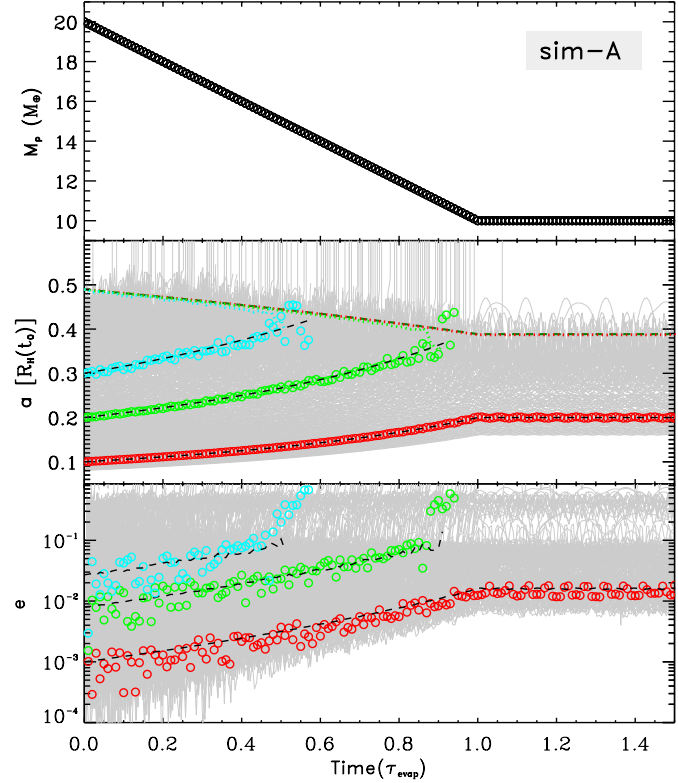}
\caption{ Evolution of mass of the planet (top), orbital semi-major axis (middle, scaled by the initial Hill radius of the planet $R_{\rm H}(t_0)$) and eccentricity (bottom) of its small moons (500 TPs, gray lines) due to photo-evaporation on the planet (Simulation A). The X axis is the evolutionary time scaled by $\tau_{\rm evap}$, where $\tau_{\rm evap}$ is the photo-evaporation time scale setting to $10^4$ times of the planet's orbital period.  Here we highlight three moons with semi-major axes started from 0.1(red), 0.2(green) and 0.3(cyan) $R_{\rm H}(t_0)$, respectively. The dotted lines in the middle panel show the critical semi-major axes for stability (Equation \ref{acr}) decrease with the planet mass.  The black dashed lines in the middle and bottom panels are theoretical predictions by Equation \ref{a_th} and Equation \ref{e_th}.}
\label{fig_A1}
\end{figure}
\end{center}

Numerically, as can be seen from Figure \ref{fig_A1}, the results are well consistent with the above analytical expectations. Three TPs start at $a_{\rm m}(t_0) =$ 0.1, 0.2 and 0.3 $R_{\rm H}(t_0)$, respectively. As the planet continues  losing mass, the three TPs increase their orbital semi-major axes and eccentricities progressively,  
while their critical semi-major axes (Equation \ref{acr}) decrease correspondingly. The two outermost TPs successively become unstable at $t \sim 0.55$ $\tau_{\rm evap}$ and $t \sim 0.95$ $\tau_{\rm evap}$ when they approach their critical semi-major axes. The innermost one survives because it is still well within its critical semi-major axis. In Figure \ref{fig_A1}, we also plot in gray lines the orbital evolutions of all (500) simulated TPs, whose final fates are counted in Table \ref{tb_fate}. About $26\%$ of the TPs survive as satellites,  the others either collide with the planet ($47\%$) or escape from orbiting the planet to become planet-like ($27\%$) bodies orbiting the star. Most survivors are those initially started within 0.2 $R_{\rm H}(t_0)$.

\subsection{Simulations B-C: small moons plus a big moon}
In this section, we consider an additional big moon added to Simulation A.  
As learned from Simulation A, the initial orbital semi-major axis of the moon plays a crucial role in determining the dynamical stability. Therefore, we study two cases in the following with the big moon starting from different semi-major axes. Furthermore, in each case, we also investigate the effect of the moon's mass by varying the mass in a large dynamical range, from $2\times10^{-10}$ $M_\earth $ to 2 $M_\earth $.
 
\subsubsection{A Stable Big Moon}
\begin{center}
\begin{figure}
\epsscale{0.95}
\plotone{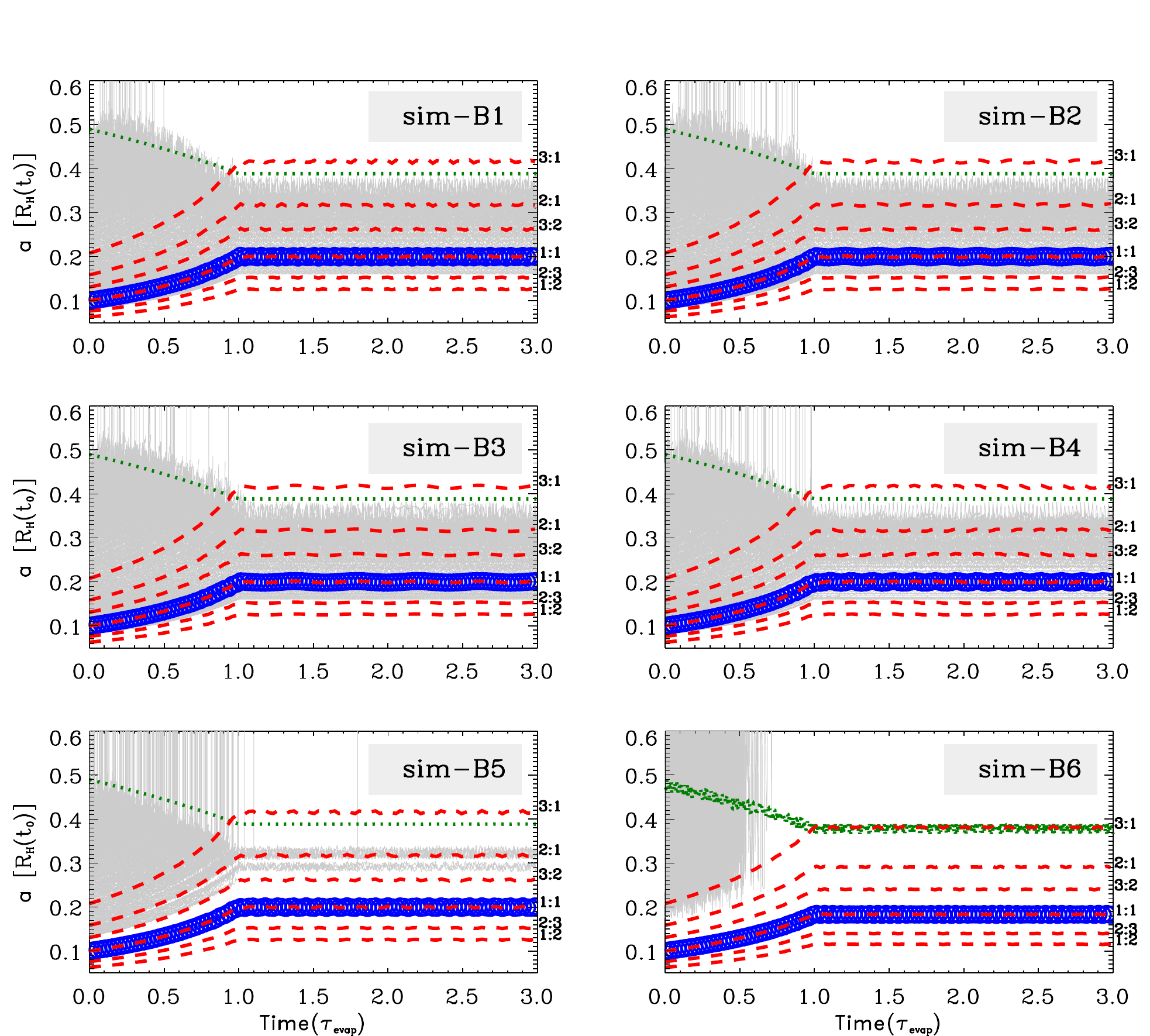}
\caption{ Evolution of moons' semi-major axes (scaled by the initial Hill radius of the planet $R_{\rm H}(t_0)$) in Simulations B1-B6. Simulation ID is presented in the top-right corner of each panel. Blue open circles represent the massive moon with an initial semi-major axis of 0.1 $R_{\rm H}(t_0)$. {\ym From simulation B1 to B6, the mass of the big moon is progressively increased by a factor of 100 from $2\times10^{-10}$ $M_\earth $ to 2 $M_\earth $. } Gray background represents the 500 test particles. Red dashed lines indicate the semi-major axes, where the orbital periods are {\ym in mean motion resonances} with the orbit of the big moon. Corresponding modes are presented on the right ordinate axis. The green dotted lines show the critical semi-major axes for stability (Equation \ref{acr}).}
\label{1p1bns_ae}
\end{figure}
\end{center}

\begin{center}
\begin{figure}
\epsscale{0.8}
\plotone{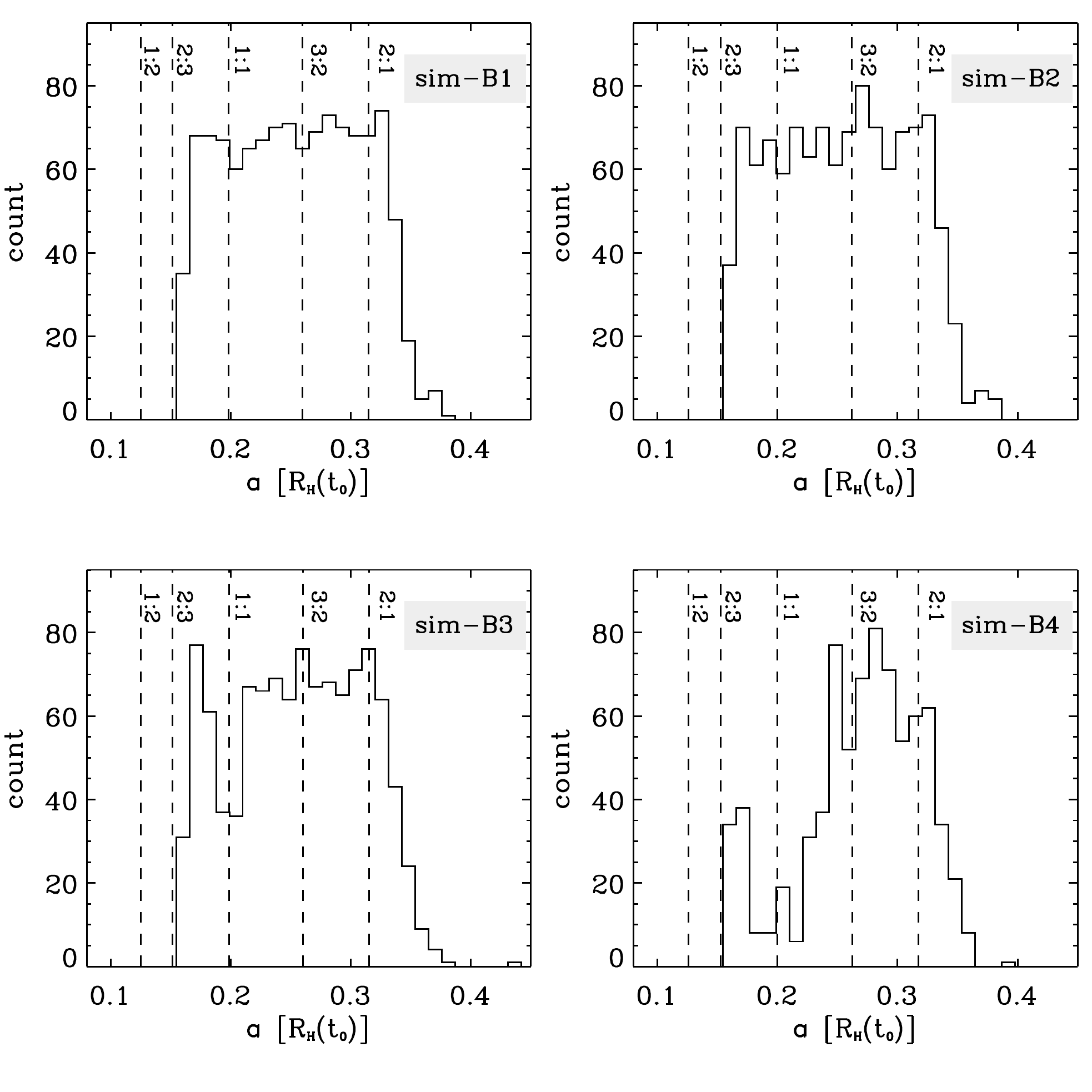}
\caption{ Distributions of semi-major axes of surviving small moons in Simulations B1-B4. Simulation ID is presented in the top-right corner of each panel. The vertical dashed lines mark the positions where the orbital periods are {\ym in mean motion resonances} with the orbit of the big moon.}
\label{1p1bns_fadist}
\end{figure}
\end{center}

In this case, the big moon is started with an initial semi-major axis of 0.1 $R_{\rm H}(t_0)$.
We study six sub-cases (B1-B6) with the mass of the big moon being progressively increased by a factor of 100 from $2\times10^{-10}$ $M_\earth $ to 2 $M_\earth $ (see Table 1). Figure \ref{1p1bns_ae} shows the evolution of semi-major axes of the moons. As can be seen, the big moon stays stable in all six simulations while the fraction of surviving small moons decreases (from 22.8\% to 0, see Table 2) as the mass of the big moon increases. Only 1.6\% of the small moons can survive as satellites if the big moon's mass is $2\times10^{-2}$ $M_\earth $. Once the big moon's mass becomes even larger, no small moons can stay orbiting around the planet. Unstable small moons have one of the following three fates: (1) hit on the surface of the planet, (2) hit the big moon, or (3) escape from the planet-satellite system to become a planet-like object orbiting the star. We note that the escape fraction (column ``Orbit star"  in Table 2) first increases (from 9.7\% to 24.8\%) as the mass of the big moons increases (from $2\times10^{-10}$ $M_\earth $ to $2\times10^{-2}$ $M_\earth $), then it significantly decreases to 3.9\% as the mass of the big moons continuously increase to 2 $M_\earth $. The turnover of the escape fraction is because the instability becomes dominated by the resonance overlap \citep{MW06} as the mass ratio of moon/planet increases to the level which is comparable to a binary (star) system. At the low moon/planet mass ratio end, resonances still affect the distribution of small moons. Figure \ref{1p1bns_fadist} plots the final semi-major axis distributions of small moons in Simulations B1-B4 (Simulations B5 and B6 are not included because almost no small moons are left at the end of simulations). As can be seen, there are similar features (gap and pileup near resonance) as compared to the asteroid belt and the planets observed by {\it Kepler} mission \citep{Fab12}. These intriguing features may be associated with overlap and asymmetry of resonances \citep{Pet13,Xie14}.

\subsubsection{An Unstable Big Moon}
\begin{center}
\begin{figure}
\epsscale{0.9}
\plotone{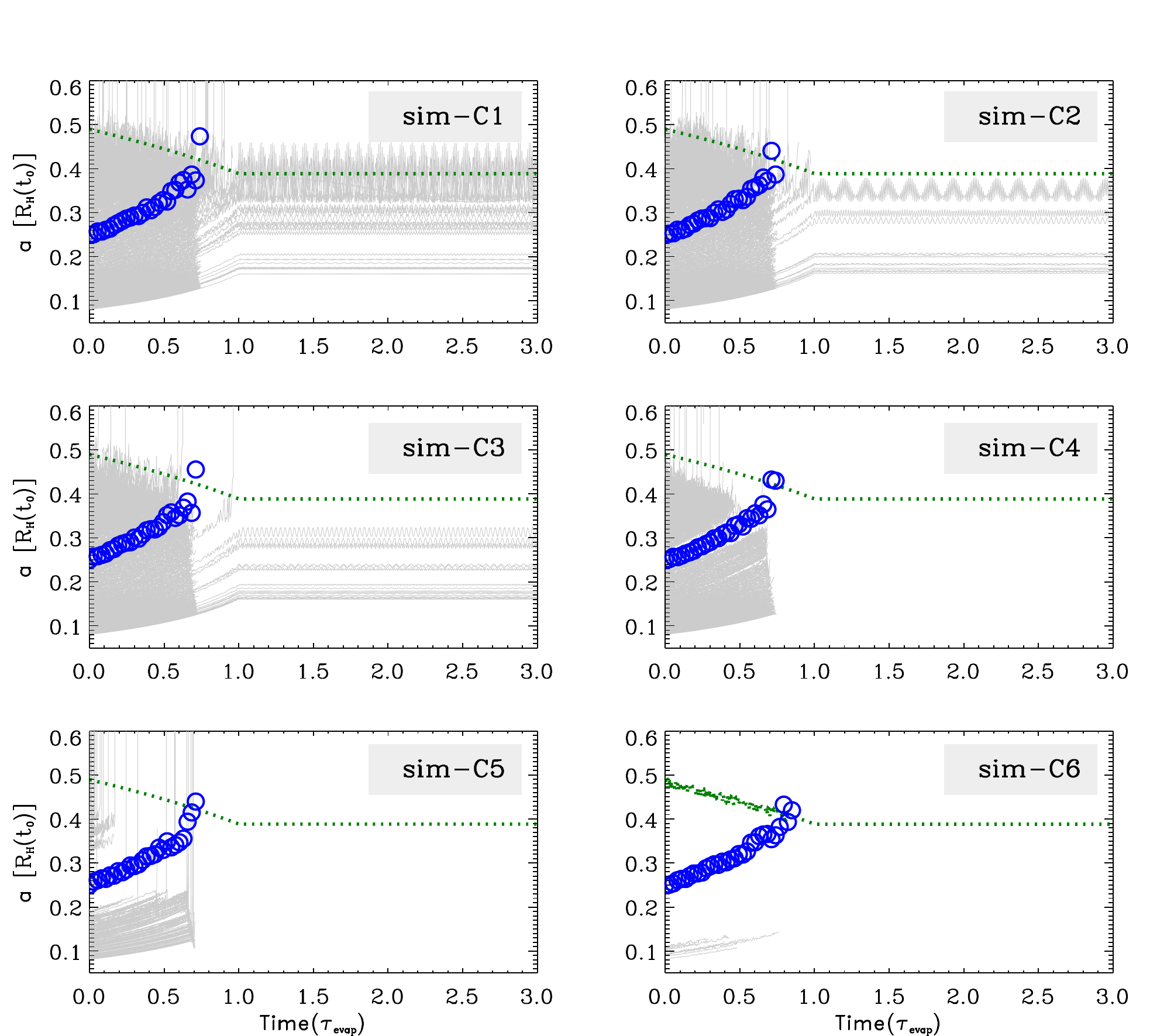}
\caption{ Evolution of moons' semi-major axes in Simulations C1-C6. Simulation ID is presented in the top-right corner of each panel. Blue open circles represent the massive moon with an initial semi-major axis of 0.25 $R_{\rm H}(t_0)$. {\ym From simulation C1 to C6, the mass of the big moon is progressively increased by a factor of 100 from $2\times10^{-10}$ $M_\earth $ to 2 $M_\earth $. } Gray background lines represent 500 test particles. Green dashed line is the theoretical stable boundary when $e \sim$ 0 (Equation \ref{alpha}). Moons become unstable when approaching this boundary.}
\label{1p1bns_ae_escape}
\end{figure}
\end{center}

In this case,  we move the big moon outwards with an initial semi-major axis of 0.25 $R_{\rm H}(t_0)$ orbiting around the planet. All other initial parameters are the same to those in Simulations B1-B6. New simulations are identified as C1-C6 in Table \ref{tb_sim} and Figure \ref{1p1bns_ae_escape}.
The evolution of orbital semi-major axes of the moons are shown in Figure \ref{1p1bns_ae_escape}. All moons' orbits expand during the photo-evaporation process. The big moon becomes unstable at $t\sim0.7$ $\tau_{\rm evap}$ when approaching the critical semi-major axis, which triggers a global instability for all the small moons.  The unstable big moon can have a very eccentric and chaotic orbit when approaching the stable boundary, thus crossing and destabilizing the orbits of small moons with a wide range of semi-major axes. A few fraction of small moons survive in Simulations C1-C3, but none can survive in Simulations C4-C6 due to the larger mass of the big moon. Compared to Simulations B, there are fewer small moons that become planet-like body orbiting the star. Most unstable small moons hit the planet or the big moon. The fraction of ``hit moon" dominates (70.4\%) in Simulation C1 and decrease to $\sim55.1\%$ in Simulation C6. This is because the increase in the mass of the big moon (from C1 to C6) enhances the ability of the big moon to scatter more small moons toward the planet, leading to an increase in the fraction of ``hit planet"  as shown in Table 2.

\subsection{Simulations D-E: All Big Moons}
In this section, we consider the full gravitational effects of all the moons, namely to set all of them as big moons in the MERCURY simulations. We study two planet-moon configurations similar to the Neptune-moon system (Simulation D) and the Uranus-moon system (Simulation E) as follows.

\subsubsection{Neptune-moon system}
For Simulation D, as in Simulations A, B and C, the planet is still assumed orbiting the star at 0.1 AU, except that, here, we set the planet-moon system as a clone of the Neptune with its seven regular moons.
Similar to previous simulations, we hypothesize the planet would lose half of its mass during the photo-evaporation process.
and study the corresponding orbital evolution of the moons. We perform 100 simulations by randomly altering the angular orbital elements of the planet and moons, e.g., the orbital argument of {\xie periapsis}, the longitude of the ascending node, and the mean anomaly. 

\begin{center}
\begin{figure}
\epsscale{0.9}
\plotone{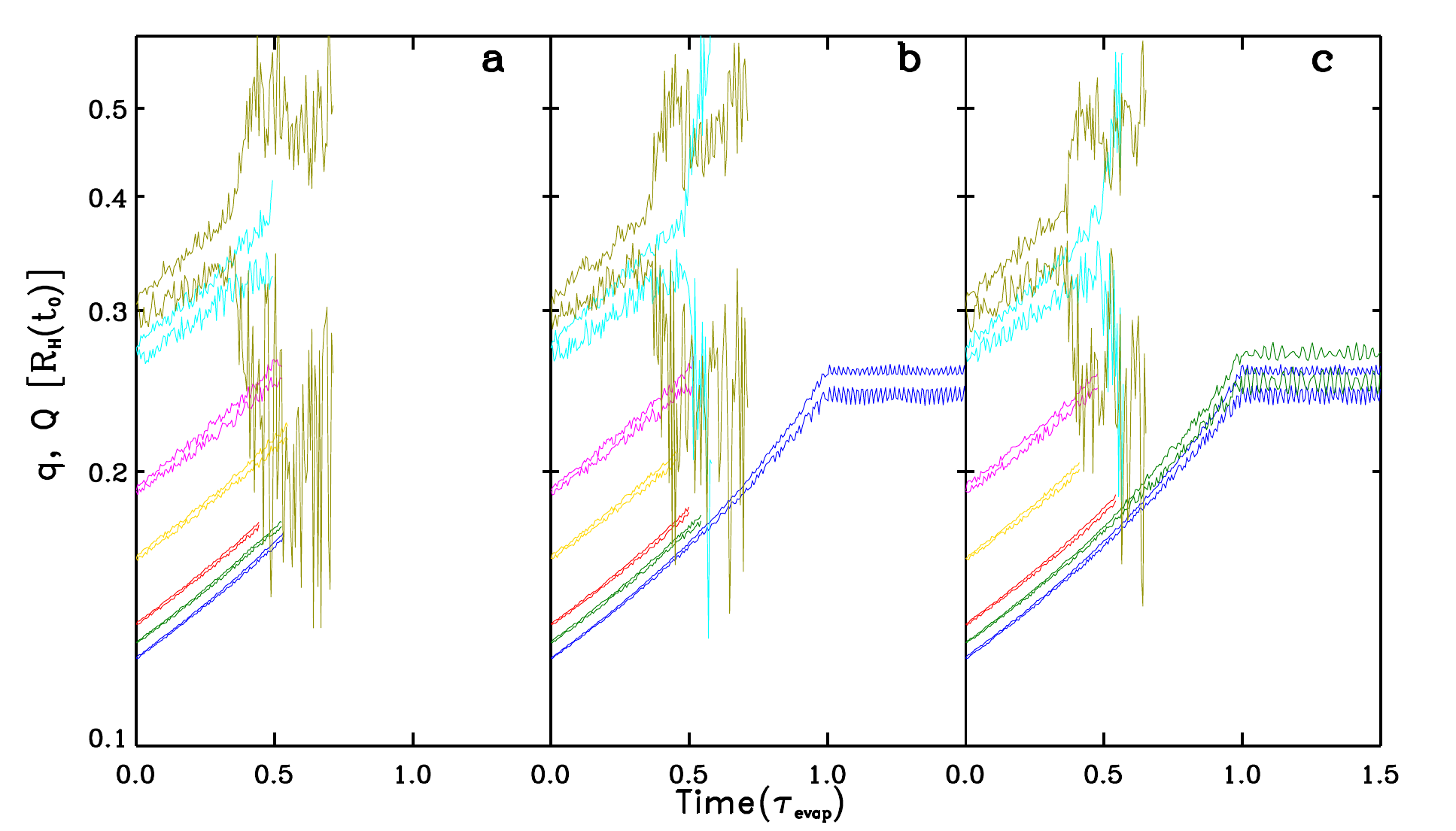}
\caption{ Orbital evolution of the moons in Simulation D. Note here the Y axis is not the semi-major axes but q (the {\xie periapsis}:$q=a_{\rm m}(1-e_{\rm m})$) and Q (the {\xie apoapsis}: $Q= a_{\rm m}(1+e_{\rm m})$) scaled by the initial Hill radius of the planet $R_{\rm H}(t_0)$. Three typical results are plotted here: panel (a) no surviving moon, panel (b) one surviving moon and panel (c) two surviving moons orbiting the planet in the end. Each moon is associated with two lines (q and Q respectively) and the same color. }
\label{neptune}
\end{figure}
\end{center}

Figure \ref{neptune} plots the results of three typical cases. Note here the Y axis is not the semi-major axes but q (the {\xie periapsis}:$q=a_{\rm m}(1-e_{\rm m})$) and Q (the {\xie apoapsis}: $Q= a_{\rm m}(1+e_{\rm m})$). The advantage of plotting q and Q here is they show the radial extension of the moon's orbit. The panel (a) of Figure \ref{neptune}, shows a case with no surviving moon. The outermost and also the most massive moon gets its orbit excited as it moves outward due to the photo-evaporation of the host planet. Its orbit crosses the inner region and hits all the other lower-mass moons around 0.5 $\tau_{\rm evap}$ and then about 0.2 $\tau_{\rm evap}$ later, the moon finally collides with the planet.
The panel (b) shows the result with one surviving moon. The outermost moon hits its four neighbor moons (except its closest neighbor---the second outermost one), then collides with the planet. The second outermost moon collides with the planet directly at $t \sim 0.6$ $\tau_{\rm evap}$.
The innermost moon is lucky to avoid all collisions and survives in the end.
The panel (c) shows the result with two surviving moons. The outermost moon hits three moons, then collides with the planet. The second innermost moon manages to survive, though hit by another lower-mass moon. The innermost moon also survives since it has no close encounter with other moons.

In all the cases, the instability of the system is driven by the orbital excitation of the outermost moon, which contributes more than 80\% of the total mass of all the moons. In most cases (87\%), the outermost moon collides with other moons and ends up with a collision with the planet. In other cases(13\%), it escapes from the planet system to become a planet-like object orbiting the star. 
In 100 simulations, we observe about 4\% simulations which end up with no moons left, similar to case (a) in Figure \ref{neptune}. For surviving moons orbiting the planet, cases (b) and (c) are most common, accounting for 42\% and 34\%, respectively.  There are also 13\%, 6\% and 1\% simulations with 3, 4 and 5 surviving moons. The innermost moon is the most-probable {\xie survivor}. In our simulations, 23.4\% of the moons still orbit the planet, 2.0\% orbit the sun, 21.2\% hit the planet, and 53.4\% hit other moons at the end of the simulation (Table \ref{tb_fate}).

\subsubsection{Uranus-moon system}
\begin{center}
\begin{figure}
\epsscale{0.9}
\plotone{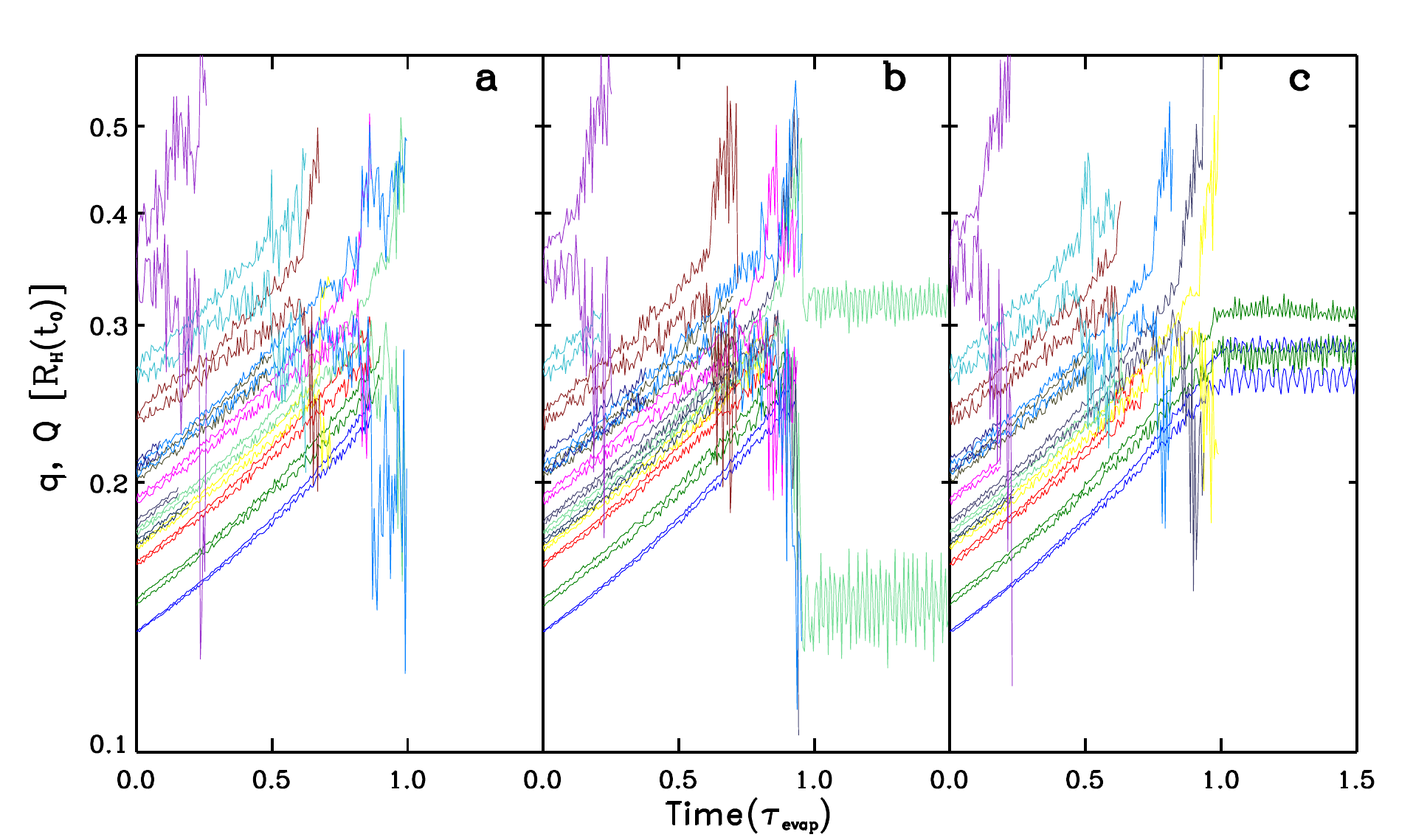}
\caption{Orbital evolution of the moons in Simulation E. Similar to Figure \ref{neptune}, except that we adopt the Uranus-moon system as the planet-moon system in simulations. }
\label{uranus}
\end{figure}
\end{center}

Simulation E is similar with Simulation D, except that we adopt the clone of the Uranus-moon system as the configuration of planet-moon system in simulations. As Uranus has many more moons,  we only consider the fourteen regular moons with semi-major axes less than 0.5 $R_{\rm H}(t_0)$. The initial conditions are presented in Table \ref{tb_sim}. Similar to Figure \ref{neptune}, we plot in Figure \ref{uranus} the results of three typical cases with 0,1 and 2 surviving moons at the end of the simulation.  The statistics of the final fates of these simulated moons are summarized in Table 2. 

Compared to Simulation D, we find the evolutions in Simulation E are generally more violent, leading to much fewer surviving moons. {\xie 9\% of the simulations end up with zero moon} (4\% in Simulation D). Most simulations (78\%) end up with only one moon, while 10\% and 3\% simulations end up with 2 and 3 moons. No system has more than 3 surviving moons.
On the other hand, the fraction of surviving moon  is 6.7\% while most moons have mutual collisions among each other (72.9\%). This is expected because Simulation E are started with a much more dynamically compact configuration with twice number of moons.

\section{DISCUSSIONS}
\subsection{Fates of Escaping Moons}
\begin{figure}
\epsscale{0.9}
\plotone{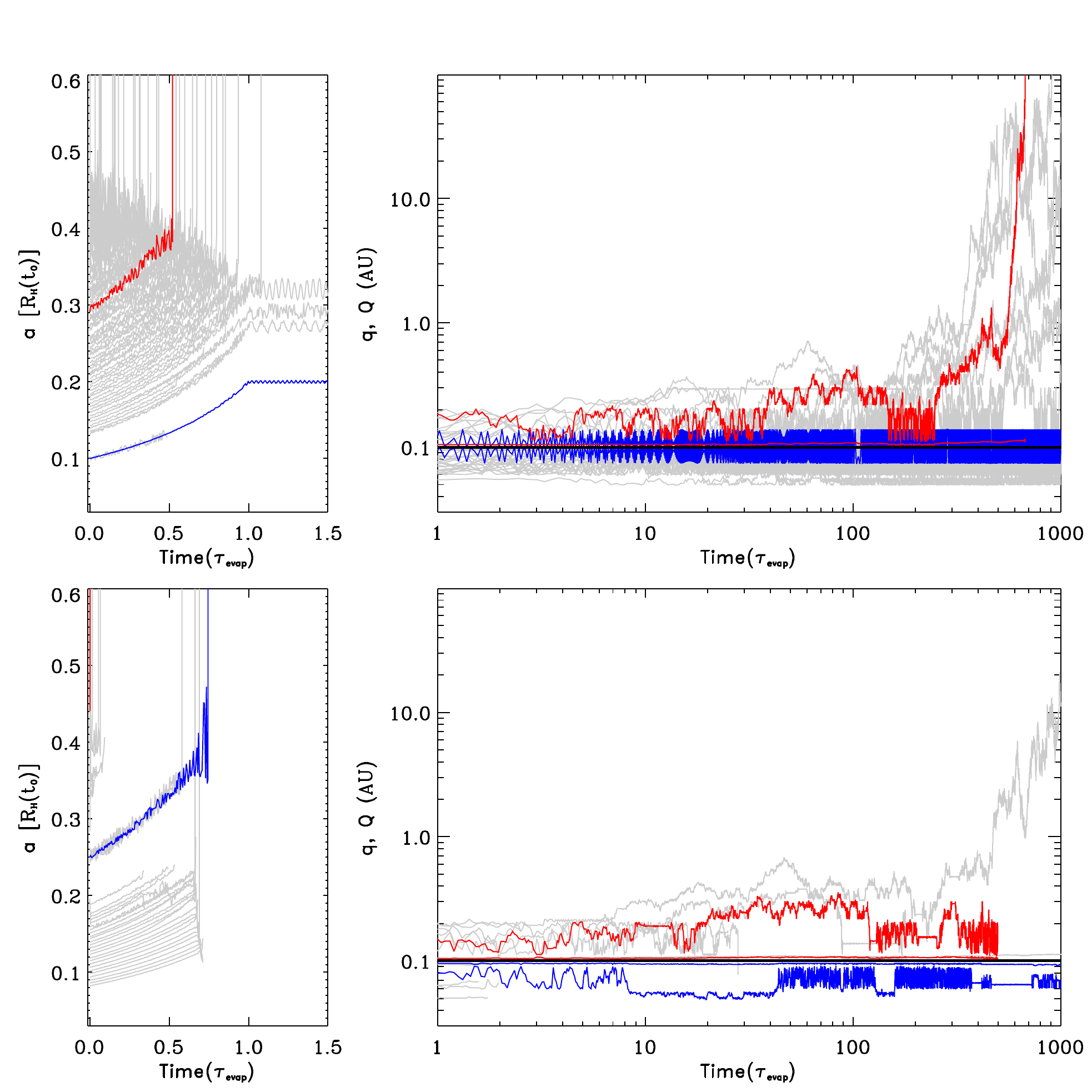}
\caption{ Long-term (1000 $\tau_{\rm evap}$) evolution of moons' orbits for Simulations B5 (top) and C5 (bottom). Left: evolution of the moons' semi-major axes with respect to the planet. Right: evolution of the moons' periastrons (q) and apstrons (Q) with respect to the center star. 
We highlight the big moon (in blue) and a certain small moon (in red) to follow their long-term evolution. Other small moons are colored in gray. The black curve in the right panel represents the semi-major axis of the planet. The unit of left panels' ordinate is the planet's initial Hill radius, and that of the right panels is the Astronomical Unit (AU). Note that each moon in the right panels is associated with two lines (q and Q respectively).}
\label{longevo}
\end{figure}

In above simulations, we note that a certain fraction (see Table 2) of moons escape from the planet-moon system to become planet-like objects orbiting the star. Here we perform long-term simulations to follow their further orbital evolutions. In order to speed up the calculation, we reduce the number of small moons to 100. The results are plotted in Figure \ref{longevo} for Simulations B5 and C5 with evolution time scale extended to 1000 $\tau_{\rm evap}$.

In Simulation B5, there are many escaping moons (24.8\%). We follow two of them which are colored in red (a small moon) and blue (the big moon). The big moon keeps orbiting the planet during the whole simulation. On the other hand, the small moon escapes at the time $\sim$0.5 $\tau_{\rm evap}$ and becomes a planet-like object orbiting the star. Afterwards, the small moon keeps its orbital periastron near the planet, but gradually increases its orbital apastron, though with large oscillations. Finally, at the time $\sim$ 700 $\tau_{\rm evap}$, the small moon boosts its apastron to beyond the boundary (100 AU) of our simulation. We speculate that the moon will be likely to escape from the star-planet system and finally to become a free-floating object in the Galaxy.

In Simulation C5, the escaping moons are fewer (8.4\%). We also follow one small moon (red) and the big moon (blue). Unlike Simulation B5, the big moon becomes a planet-like object orbiting the star at the time $\sim$0.7 $\tau_{\rm evap}$. Then it keeps its orbital apastron near the planet but with its periastron oscillating between 0.05 and 0.1 AU. The moon would have close encounters with the planet near its periastron and has the possibility to hit the planet. However, interactions with other planets or planetesimals (not included in our current simulations) may further change its orbit to let it become a stable planet. On the other hand, the small moon (red) finally collides with the planet at the time $\sim$ 500 $\tau_{\rm evap}$. 

As learned from above long-term simulations, we see that the fates of those escaping moons are diverse. They can further escape from the star-planet system to become rogue planets in the Galaxy or become stable planets orbiting the star or come back to collide with their parent planet after a certain time orbiting the star. 

{\xie
\subsection{Effects of other Model Parameters}
In our simulations shown above, we simply adopt a linear photo-evaporation model with a time scale of $\tau_{\rm evap}=10^4$ $P_{\rm p}$. In reality, the process of photo-evaporation is much longer ($10^7-10^8$ yr $\sim 10^9$ $P_{\rm p}$) and definitely nonlinear with time. In order to study the effects of these photo-evaporation model parameters, we performed more simulations with linear/non-linear mass loss process of various time scales. The results are plotted in Figure 8 and Figure 9.  As can be seen, regardless of whether linear or non-linear mode, the dynamical results are similar as long as the photo-evaporation time scale is much longer than the planet's orbital period, i.e,. $\tau_{\rm evap}>(10^2-10^3)$ $P_{\rm p}$. This is expected because the evolution approaches an adiabatic progress for such a large $\tau_{\rm evap}$. Therefore, we choose $\tau_{\rm evap}=10^4$ $P_{\rm p}$ in our previous simulations to both meet the adiabatic condition and save the computation time. 

In our nominal models, we simply fix the planet density (1.66 $g/cm^3$, same as Neptune). In reality, as the planet loses mass due to photo-evaporation, its density and radius may change, which can modify the ability of the moon to hit the planet. In order to quantify these effects, we perform another two simulations. Both simulations have the same initial conditions as simulation A, except that the planetary densities are a factor of two larger and smaller, respectively. We find the results of these two simulations are comparable to that of simulation A. The fraction of surviving moons does not change at all (both are $26\%$ same to simulation A). The only difference is the fraction of moons that hit the planet, which increases (decreases) from 47 to 55\% (30\%) for the simulation with smaller (larger) planet density.  The results are expected as lower density leads to larger planetary radius {\ym for a given planetary mass} and thus larger cross-section for planet-moon collision. The stability of the moon is determined by the Hill radius of the planet (independent of planet radius), thus there is no change in surviving moons if only the planet radius (density) varies. Therefore, we conclude that changing the planetary density (thus radius) would not qualitatively change our major result, namely, photo-evaporation plays a destructive role in the orbital evolution of the moon system, generally leading to global instability of exo-moon system.
}

\subsection{Implications to Observations} 
The {\it Kepler} mission has detected more than 4600 exoplanets/candidates. A large portion ($\sim$40\%) of them are in close-in orbits with orbital semi-major axes $<$ 0.1 AU, for which photo-evaporation by the host star could be relevant. Indeed, \cite{OW13} find that about 50\% of {\it Kepler} planet candidates may have been significantly eroded by photo-evaporation. If those planets had moons, many of these moons would have been lost due to the instability triggered by the photo-evaporation as shown in above simulations. As photo-evaporation depends on the proximity to the host star, we expect a gradient in observation: the occurrence rate of exomoon significantly drops as approaching ($<$0.1 AU) the center star. Furthermore, the X-ray exposure plays an important role in photo-evaporation and the X-ray flux varies greatly for different stars, with late-type stars being significantly more X-ray luminous \citep{Gud04, Jac12}. Therefore, we expect that the instability induced by photo-evaporation also has a dependency on the spectrum type of the host star. From this aspect, we predict that exomoons are less likely to be found around M stars as compared to earlier type (F, G, and K) stars.

\section{SUMMARY}
Many planet candidates found by {\it Kepler} mission are in close orbits around their host stars, whose photon radiation could evaporate the atmosphere of the close-in planets. In this paper, we model this photo-evaporation process as an adiabatic mass loss on the planet, and numerically investigate the corresponding effects on the dynamical evolution of the moons' orbits around the planet.  

We begin with the simplest case, where the moons are treated as test particles (small moons) without considering their mutual gravity (Section 2.1). Simulation A illustrates the direct effects of photo-evaporation on the moons, namely expanding (increase $a_{\rm m}$) and exciting  (increase $e_{\rm m}$) the moons' orbits (Figure 1) , which triggers orbital instability as moons' orbits cross the boundary of stability (Equation 1).  Next, we consider a set of new simulations (Simulations B and C) by adding a big moon with various masses and initial semi-major axes. These simulations help us understand the role of the moon's gravity in developing the orbital instability. As expected, a moon with greater mass and larger initial semi-major axis (thus closer to the stability boundary) tend to make the system more unstable, with almost all moons being lost in the end (Figures 2, 3 and 4). Finally, we perform two more realistic simulations (Simulations D and E) by cloning the moon systems of Neptune and Uranus (Figures 5 and 6). {\ym In these two set of simulations, mutual gravity of all moons have been fully considered, which leads to more chaotic evolution of the systems.}

In any case, we learn that photo-evaporation plays a destructive role in the orbital evolution of the moon system, generally leading to moon loss. The fates of the lost moons are diverse. While the majority of them are likely to collide with other moons or with the planet, some of them could escape from the moon system to become a new planet orbiting the star or even a free-floating object in the Galaxy (Figure 7). Based on our simulations, we therefore speculate that exomoons are fewer around planets that are close ($<0.1$ AU) to their host stars, especially M type stars, because they are more X-ray luminous and thus enhancing photo-evaporation. On-going or future exomoon searching programs should consider above effects for their target selection. 

\acknowledgments
This research is supported by the Key Development Program of Basic Research of China (973 program, Grant No. 2013CB834900), the National Natural Science Foundation of China (Grant No. 11003010, 11333002, 11403012 and 11503009), a Foundation for the Author of National Excellent Doctoral Dissertation (FANEDD) of PR China,  the Strategic Priority Research Program "The Emergence of Cosmological Structures" of the Chinese Academy of Sciences (Grant No. XDB09000000) and the Natural Science Foundation for the Youth of Jiangsu Province (No. BK20130547).
\clearpage



\begin{figure}
\epsscale{0.9}
\plotone{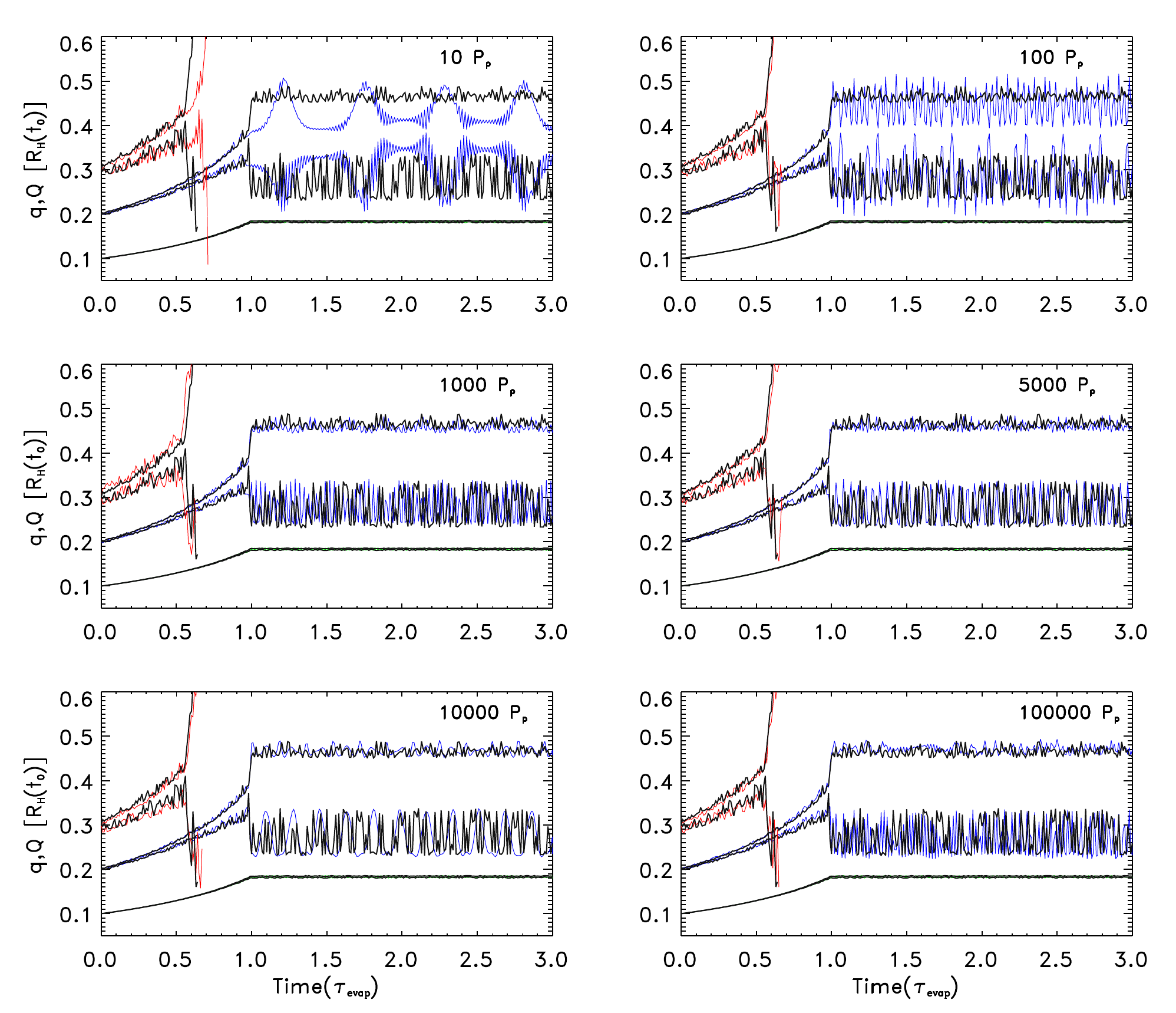}
\caption{ {\xie Orbital evolutions with different $\tau_{\rm evap}$. The time scale is shown in the top-right corner of each panel. The Y axis is q (the periapsis: $q=a_{\rm m}(1-e_{\rm m})$) and Q (the apoapsis: $Q= a_{\rm m}(1+e_{\rm m})$) scaled by the initial Hill radius of the planet $R_{\rm H}(t_0)$. Here we plot three moons with semi-major axes started from 0.1(green), 0.2(blue) and 0.3(red) $R_{\rm H}(t_0)$, respectively. Each moon is associated with two lines (q and Q respectively). For better comparison, in each panel, we plot simulation results (black lines) with the longest time scale ($\tau_{\rm evap}=10^6 ~ P_{\rm p}$). We see that the results are almost identical if the photo-evaporation time scale is sufficiently large, e.g., $\tau_{\rm evap}>$1000 P$_{\rm p}$. }} 
\label{test.timescale}
\end{figure}

\begin{figure}
\epsscale{0.9}
\plotone{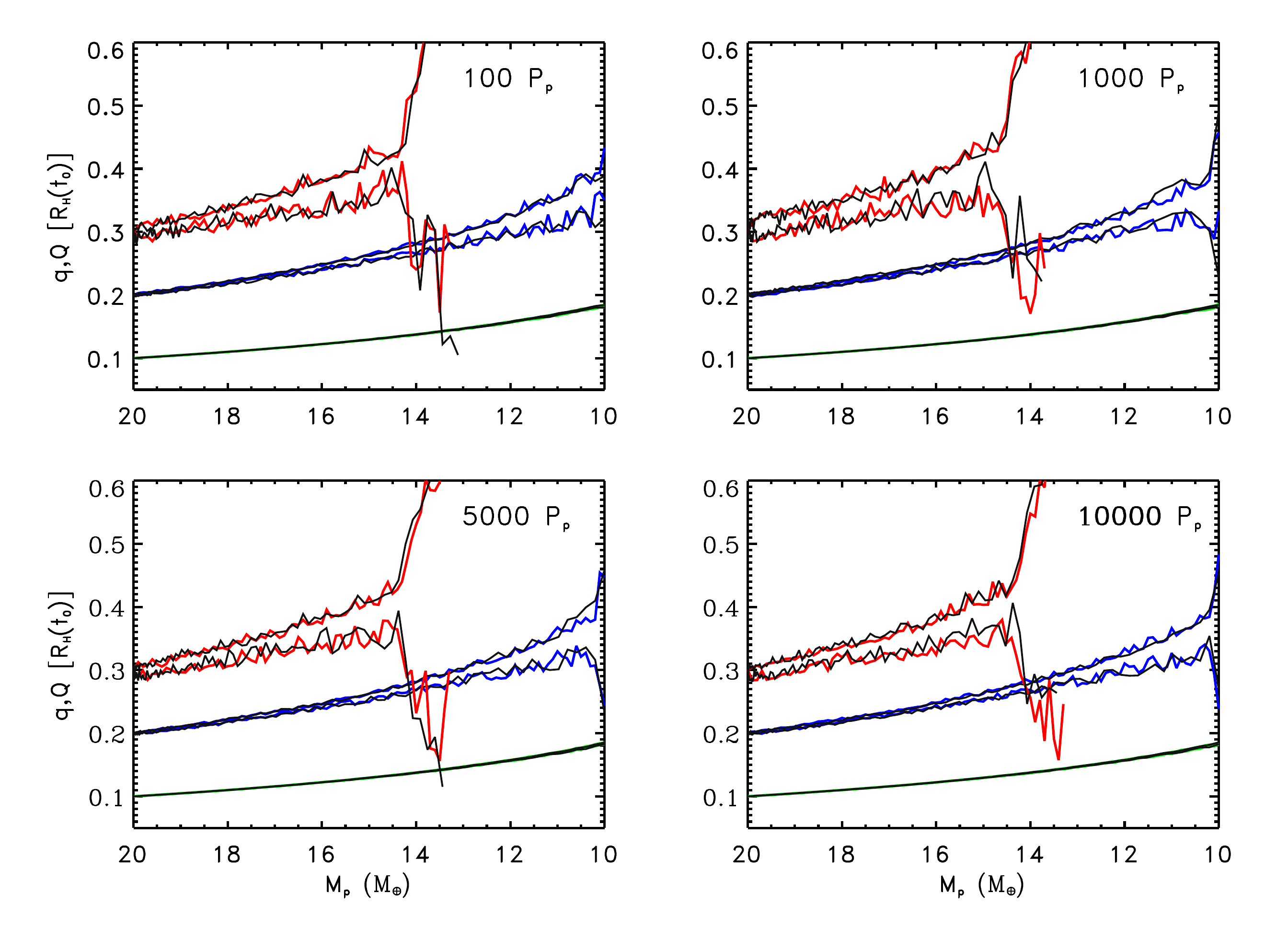}
\caption{ \xie Orbital evolutions with different $\tau_{\rm evap}$ and different planet mass loss mode (linear: $M_{\rm p}(t)/M_{\rm p}(t_0) = 1 - {1\over 2} \left(t / \tau_{\rm evap}\right)$ (black lines); nonlinear: $M_{\rm p}(t)/M_{\rm p}(t_0) = 1 - {1\over 2} \left(t / \tau_{\rm evap}\right)^2$ (red, blue and green lines)). Note that the X axes are the mass of the planet (NOT time). The Y axes are the same to those in Figure 8.} 
\label{test.timescale}
\end{figure}

\clearpage
\newpage
\renewcommand\arraystretch{1.2}
\begin{deluxetable}{c ccc cc ccc}
\tabletypesize{\scriptsize}
\tablewidth{0pt}
\tablecaption{ Initial Conditions of Simulations \label{tb_sim}} 
\tablehead{
\colhead{\multirow{3}{*}{Simulation}} 	& 	\multicolumn{3}{c}{Planet} 	&
\multicolumn{2}{c}{Test Particle (TP, small Moon)} 		&	\multicolumn{3}{c}{Big Moon} \\
\cmidrule(r){2-4} \cmidrule(r){5-6}  \cmidrule(r){7-9} 		&
\colhead{$M_{\rm pi}$}	&  	\colhead{$M_{\rm pf}$} 		&  	
\colhead{$a_{\rm p}$}   	& 	\colhead{$N_{\rm t}$}  		& 	
\colhead{$a_{\rm t}$}    	& 	\colhead{$N_{\rm m}$}  		&	
\colhead{$M_{\rm m}$}  	&	\colhead{$a_{\rm m}$}     	\\
\colhead{}     				&	\colhead{($M_\earth$)} 	&
\colhead{($M_\earth$)} 		&   	\colhead{(AU)}   			& 
\colhead{}   					& 	\colhead{($R_{\rm H}(t_0)$)}   & 
\colhead{}  					& 	 \colhead{($M_\earth$)}  	& 
\colhead{($R_{\rm H}(t_0)$)} 
}
\startdata
\multirow{2}{*}{A}	&	\multirow{2}{*}{20}	&	\multirow{2}{*}{10}	&	\multirow{2}{*}{0.1}	&	\multirow{2}{*}{500}	&	\multirow{2}{*}{0.08 $\sim$ 0.49}				&	\multirow{2}{*}{0}		&	\multirow{2}{*}{\nodata}	&	\multirow{2}{*}{\nodata}	\\
	&	&	&	&	&	&	&	&		\\
\hline
\multirow{2}{*}{B1-B6}	&	\multirow{2}{*}{20}	&	\multirow{2}{*}{10}	&	\multirow{2}{*}{0.1}	&	\multirow{2}{*}{500}	&	\multirow{2}{*}{0.08 $\sim$ 0.49}				&	\multirow{2}{*}{1}		&	\multirow{2}{*}{$2 \times 10^{(-10, -8, -6, -4, -2, 0)}$}	&	\multirow{2}{*}{0.1}	\\
	&	&	&	&	&	&	&	&		\\	
\hline
\multirow{2}{*}{C1-C6}	&	\multirow{2}{*}{20}	&	\multirow{2}{*}{10}	&	\multirow{2}{*}{0.1}	&	\multirow{2}{*}{500}	&	\multirow{2}{*}{0.08 $\sim$ 0.49}				&	\multirow{2}{*}{1}		&	\multirow{2}{*}{$2 \times 10^{(-10, -8, -6, -4, -2, 0)}$}	&	\multirow{2}{*}{0.25}	\\
	&	&	&	&	&	&	&	&		\\
\hline	
\multirow{2}{*}{D}	&	\multirow{2}{*}{17.2}	&	\multirow{2}{*}{8.6}	&	\multirow{2}{*}{0.1}	&	\multirow{2}{*}{\nodata}	&	\multirow{2}{*}{\nodata}		&	\multirow{2}{*}{7}	&	\multirow{2}{*}{$8.4 \times 10^{-10} \sim 8.4 \times 10^{-6}$}	&	\multirow{2}{*}{0.13 $\sim$ 0.30}	\\
	&	&	&	&	&	&	&	&		\\	
\hline
\multirow{2}{*}{E}	&	\multirow{2}{*}{14.6}	&	\multirow{2}{*}{7.3}	&	\multirow{2}{*}{0.1}	&	\multirow{2}{*}{\nodata}	&	\multirow{2}{*}{\nodata}		&	\multirow{2}{*}{14}	&	\multirow{2}{*}{$6.4 \times 10^{-10} \sim 1.1 \times 10^{-5}$	}&	\multirow{2}{*}{0.14 $\sim$ 0.36}	\\
	&	&	&	&	&	&	&	&		\\	
\enddata
\tablenotetext{a}{$M_{\rm pi}$, $M_{\rm pf}$ and $a_{\rm p}$ are planet's mass before and after photo-evaporation, and the orbital semi-major axis around the central star, respectively. $N_{\rm t}$ and $a_{\rm t}$ are the number and semi-major axes of test particles around the planet. $N_{\rm m}$, $M_{\rm m}$ and $a_{\rm m}$ are the number, mass, semi-major axes of moons around the planet, respectively. $R_{\rm H}(t_0)$ is the initial Hill radius of the planet. }
\tablenotetext{b}{For clarify, some common initial parameters are not listed here. For example, we consider only one planet orbiting a central star with mass of 1 $M_\odot$. The evaporation time scale is set as $10^4$ $P_{\rm p}$, where $P_{\rm p}=365.25\times(0.1/1 \rm AU)^{1.5}\sim11.6$ d, is the orbital period of the planet.}
\end{deluxetable}
\renewcommand\arraystretch{1.0}

\renewcommand\arraystretch{1.2}
\begin{deluxetable}{ccccc}
\tabletypesize{\small}
\tablewidth{0pt}
\tablecaption{ Fates of simulated Moons  \label{tb_fate}}
\tablehead{
\colhead{Simulation} & \colhead{Orbit planet} & \colhead{Orbit star} & \colhead{Hit planet} & \colhead{Hit moon} \\
\colhead{} & \colhead{(\%)} & \colhead{(\%)} & \colhead{(\%)} & \colhead{(\%)}
}
\startdata
A & 26.0 & 27.0 & 47.0 & \nodata \\
\hline
B1	&	22.8	&	9.7	&	20.4	&	47.1	\\
B2	&	22.7	&	15.8	&	32.4	&	29.1	\\
B3	&	21.5	&	13.7	&	28.9	&	35.9	\\
B4	&	15.4	&	17.0	&	28.1	&	39.5	\\
B5	&	1.6	&	24.8	&	40.2	&	33.4	\\
B6	&	0.0	&	3.9	&	52.6	&	43.5	\\
\hline
C1	&	3.3	&	8.8	&	17.5	&	70.4	\\
C2	&	3.4	&	3.3	&	13.0	&	80.3	\\
C3	&	2.1	&	3.3	&	8.0	&	86.6	\\
C4	&	0.0	&	3.9	&	7.2	&	88.9	\\
C5	&	0.0	&	8.4	&	24.4	&	67.2	\\
C6	&	0.0	&	0.5	&	44.4	&	55.1	\\
\hline
D  &  	23.4  &  2.0  &  	21.2  	&  53.4  	\\
\hline
E  &  	6.7  &  0.9  	&  	19.5  	&  72.9  	\\
\enddata
\tablenotetext{a}{For Simulations A, B1-B6 and C1-B6, we give the statistics of test particles' fates. For Simulations D and E, whose moons are all massive, we give the statistics of all the massive moons' fates. Column ``Orbit planet" gives the fraction of moons that finally survive as orbiting the planet. Column ``Orbit star"  gives the fraction of moons that finally escape from the planet-moon system and become planet-like objects orbiting the star. Columns ``Hit planet" and ``Hit moon" give the fraction of moons that finally collide with the planet and other (big) moons, respectively. }
\end{deluxetable}
\renewcommand\arraystretch{1.0}

\end{document}